\documentclass[conference]{IEEEtran}
\IEEEoverridecommandlockouts
\usepackage[pdftex]{graphicx}
\usepackage{amsmath,amsfonts,amssymb}
\usepackage{mathtools}
\usepackage{bm}
\usepackage{cite}
\usepackage[ruled,vlined]{algorithm2e}
\SetAlCapNameFnt{\small}
\SetAlCapFnt{\small}
\usepackage{siunitx}
\usepackage{mleftright}
\usepackage{amsthm}
\newtheorem{theo}{Theorem}
\newtheorem{prop}{Proposition}

\theoremstyle{definition}
\newtheorem{exmp}{Example}

\newtheorem{rmrk}{Remark}

\theoremstyle{remark}

\usepackage{setspace}
\usepackage[caption=false,font=footnotesize]{subfig}
\DeclareMathOperator*{\minimize}{minimize}

\usepackage{textcomp}
\usepackage{xcolor}
\usepackage{flushend}
\begin{document}

\title{Theoretical Validation of the Latent Optimally Partitioned-$\ell_2/\ell_1$ Penalty with Application to Angular Power Spectrum Estimation%
\thanks{\scriptsize The first author conducted part of this work as a guest researcher at Fraunhofer Heinrich Hertz Institute.
The first and third authors were supported by JSPS Grants-in-Aid under Grants 21K17827, 23K22762, and 26K17380.
The second author acknowledges the financial support from the
Federal Ministry of Research, Technology and Space (BMFTR) in Germany,
project xG-RIC (grants 16KIS2429K and 16KIS243), and from the 6G-MIRAI
project, which has received funding from the Smart Networks and
Services Joint Undertaking (SNS JU) under the European Union's Horizon
Europe research and innovation program under Grant Agreement No
10119236. Views and opinions expressed are, however, those of the
author(s) only, and they do not necessarily reflect those of the
European Union or the SNS JU (granting authority). Neither the European
Union nor the granting authority can be held responsible for them.}\vspace{-2pt}}

\author{\IEEEauthorblockN{Hiroki Kuroda}
\IEEEauthorblockA{\textit{Institute of Science Tokyo}\\
kuroda@ict.eng.isct.ac.jp}\vspace{-10pt}
\and
\IEEEauthorblockN{Renato~L.~G.~Cavalcante}
\IEEEauthorblockA{\textit{Fraunhofer Heinrich Hertz Institute}\\
renato.cavalcante@hhi.fraunhofer.de}\vspace{-10pt}
\and
\IEEEauthorblockN{Masahiro~Yukawa}
\IEEEauthorblockA{\textit{Keio University}\\
yukawa@elec.keio.ac.jp}\vspace{-10pt}}

\maketitle
\setlength{\abovedisplayskip}{5pt}
\setlength{\belowdisplayskip}{5pt}

\begin{abstract}
This paper demonstrates that, in both theory and practice, the latent optimally partitioned (LOP)-$\ell_2/\ell_1$ penalty is effective for exploiting block-sparsity without knowledge of the concrete block structure. More precisely, we first present a novel theoretical result showing that the optimized block partition in the LOP-$\ell_2/\ell_1$ penalty satisfies a condition required for accurate recovery of block-sparse signals. Motivated by this result, we present a new application of the LOP-$\ell_2/\ell_1$ penalty to estimation of angular power spectrum, which is block-sparse with unknown block partition, in MIMO communication systems. Numerical simulations show that the proposed use of block-sparsity with the LOP-$\ell_2/\ell_1$ penalty significantly improves the estimation accuracy of the angular power spectrum.
\end{abstract}

\section{Introduction}
In many applications of signal processing and machine learning,
target signals to be estimated are block-sparse, i.e., their nonzero components are clustered in a few blocks.
Earlier studies
demonstrate the effectiveness of the mixed $\ell_2/\ell_1$ norm
as a penalty function for a block-sparse signal with a \emph{known} block partition \cite{Stojnic:BlockSparse,Eldar:GroupSparse,Lv:GroupSparse,Elhamifar:GroupSparse,Yuan:l12,Nardi:BlockSpare,Wei:BlockSparse,Huang:Review}.
However,
the information of block partition is not available in many applications, e.g.,
audio and video processing \cite{Yu:Audio,Cevher:Video},
change detection \cite{Kuroda_FPC},
and remote sensing \cite{Kitahara:PAWR_GRSS}.

To cope with the problem of unknown block partition,
the study in \cite{Kuroda:BlockSparse} proposes
the latent optimally partitioned (LOP)-$\ell_2/\ell_1$ penalty
as a tight convex relaxation of the combinatorial penalty
that minimizes
the mixed $\ell_2/\ell_1$ norm over the set of candidate block partitions.
Although excellent performance of the LOP-$\ell_{2}/\ell_{1}$ penalty 
has been experimentally demonstrated \cite{Kuroda:BlockSparse,Kuroda:GraphSparse,Kuroda:PSDEstimation,Yu:l0LOP,Furuhashi:LOP_image},
so far it has not been revealed whether
the block partition optimized by the LOP-$\ell_{2}/\ell_{1}$ penalty
becomes the desired block partition required for accurate recovery of the target signal.

The first objective of this paper is to present a novel theoretical result demonstrating that
the LOP-$\ell_{2}/\ell_{1}$ penalty indeed selects the desired block partition.
As discussed in Sect.~\ref{sect:desiredBlockl2l1},
for accurate recovery of a block-sparse signal,
the block partition is desired to separate
components of (widely) different magnitudes.
Our main result (Theorem \ref{theo:ConditionLOPpenEqualL2L1})
demonstrates that
the block partition optimized by
the LOP-$\ell_{2}/\ell_{1}$ penalty satisfies this desired property
if its tuning parameter controlling the number of blocks is suitably selected.

The second objecctive of this paper is to present a novel application
of the LOP-$\ell_{2}/\ell_{1}$ penalty to angular power spectrum (APS) estimation, which
is a challenging problem
with important applications
in MIMO communication systems
\cite{Decurninge:LocalizationMIMO,Khalilsarai:NNLS_APS,Miretti:CovarianceConversion,Bameri:CovarianceConversion,Kaneko:APS_AdaptGaussKernel,Cavalcante:PilotDecont}.
Many studies indicate that
the APS is block-sparse in practical propagation environments,
but the block-sparsity of the APS has been considered  difficult to exploit
because
the desired block partition is
unknown owing to its dependence on the propagation environment \cite{Song:ML_APS_GroupSparse,Khalilsarai:ML_APS_GroupSparse}.
Motivated by our theoretical result,
we attempt to resolve this difficulty by integrating the LOP-$\ell_{2}/\ell_{1}$ penalty
into an optimization problem in \cite{Cavalcante:ModelDataAPS} that considers the physical observation model and the dataset of past APS.
Numerical simulations based on the 3GPP technical document \cite{TechRep:3gpp}
demonstrate that the estimation accuracy of the APS is significantly
improved with the proposed use of block-sparsity.

\section{Preliminaries}
Let $\mathbb{R}$, $\mathbb{R}_{+}$, $\mathbb{R}_{++}$, and $\mathbb{C}$
denote the sets of
real numbers, nonnegative real numbers, positive real numbers, and complex numbers, respectively.
The cardinality of a set
$S$ is denoted by $|S|$.
We denote the transpose and the Hermitian transpose by $(\cdot)^{\mathsf{T}}$
and $(\cdot)^{\mathsf{H}}$, respectively.
We define the support of $\bm{x} \in \mathbb{R}^{N}$ by
$\mathrm{supp}(\bm{x}) := \{n \in \{1,\ldots,N\}\mid x_n \neq 0\}$.
The $\ell_1$ norm, the $\ell_2$ norm, and the $\ell_\infty$ norm
of $\bm{x} \in \mathbb{R}^{N}$ are denoted by
$\|\bm{x}\|_1 := \sum_{n=1}^{N}|x_n|$,
$\|\bm{x}\|_2 := \sqrt{\bm{x}^{\mathsf{T}}\bm{x}}$, and
$\|\bm{x}\|_\infty := \max_{n \in \{1,\ldots,N\}}|x_n|$, respectively.
We define the $\ell_{\infty}$ operator norm of $\bm{A} \in \mathbb{R}^{M \times N}$ by
$\|\bm{A}\|_{\infty} := \max \{\|\bm{A}\bm{x}\|_{\infty} \mid \bm{x} \in \mathbb{R}^{N}, \|\bm{x}\|_{\infty} \leq 1\}$.
For $\bm{x}\in\mathbb{R}^N$ and $\mathcal{I} \subset \{1,\ldots,N\}$,
the subvector of 
$\bm{x}$ indexed by $\mathcal{I}$ is denoted by
$\bm{x}_{\mathcal{I}} :=  (x_n)_{n \in \mathcal{I}} \in \mathbb{R}^{|\mathcal{I}|}$.
For $\bm{A} = (a_{m,n}) \in\mathbb{R}^{M \times N}$ and index sets $\mathcal{I} \subset \{1,\ldots,M\}$ and $\mathcal{J} \subset \{1,\ldots,N\}$, we denote the submatrix of $\bm{A}$
indexed by $\mathcal{I}$ and $\mathcal{J}$ by
$\bm{A}_{\mathcal{I}, \mathcal{J}} := (a_{m,n})_{(m,n) \in \mathcal{I}\times\mathcal{J}}
\in \mathbb{R}^{|\mathcal{I}|\times|\mathcal{J}|}$.
\subsection{Desired block partition for the mixed $\ell_2/\ell_1$ norm}
\label{sect:desiredBlockl2l1}
We say that $(\mathcal{B}_k)_{k=1}^{K}$ is a block partition of $\{1,\ldots,N\}$ if
$(\mathcal{B}_k)_{k=1}^{K}$ is a partition of $\{1,\ldots,N\}$
and there exist $(n_k, m_k) \in \{1,\ldots,N \}^2$ such that
$\mathcal{B}_k = \{n \in \{1,\ldots,N\}\mid n_{k} \leq n \leq m_k \}$ for every $k \in \{1,\ldots,K\}$.
Suppressing the value of the mixed $\ell_2/\ell_1$ norm
\begin{align}
\label{eq:def:mixedl2l1norm}
\|\bm{x} \|_{2,1}^{(\mathcal{B}_k)_{k=1}^{K}} := \sum_{k=1}^{K}\sqrt{|\mathcal{B}_k|}\|\bm{x}_{\mathcal{B}_k} \|_2
\end{align}
promotes
block-sparsity of $\bm{x} \in \mathbb{R}^{N}$ with the block partition $(\mathcal{B}_k)_{k=1}^{K}$
by pushing components in $\mathcal{B}_k$ zeros or nonzeros together.
It should be noted that the $\ell_2$ norm $\|\bm{x}_{\mathcal{B}_k} \|_2$
penalizes $x_n$ for some $n \in \mathcal{B}_k$ heavily
if the magnitude of $x_n$ is larger than other components in $\mathcal{B}_k$.
This property enables to promote the block-sparsity,
but also causes estimation error if magnitudes of components in the same block
are unbalanced.
For example, suppose that a block $\mathcal{B} \subset \{1,\ldots,N\}$ is decomposed
as $\mathcal{B} = \mathcal{B}' \cup \mathcal{B}''$, and
the magnitudes of $(x_n)_{n \in \mathcal{B}'}$ are significantly larger
than those of $(x_n)_{n \in \mathcal{B}''}$.
Then, since $\|\bm{x}_\mathcal{B}\|_2 = \sqrt{\sum_{n\in \mathcal{B}'}x_n^2 +  \sum_{n\in \mathcal{B}''}x_n^2}$
is dominated by $\sum_{n\in \mathcal{B}'}x_n^2$,
suppressing the value of $\sqrt{|\mathcal{B}|}\|\bm{x}_\mathcal{B}\|_2$
causes severe underestimation of the magnitudes of $(x_n)_{n \in \mathcal{B}'}$.
Since $|\mathcal{B}| > |\mathcal{B}'|$,
this underestimation effect is more severe than suppressing
$\sqrt{|\mathcal{B}'|}\|\bm{x}_{\mathcal{B}'}\|_2+\sqrt{|\mathcal{B}''|}\|\bm{x}_{\mathcal{B}''}\|_2$.
Note that the weight $\sqrt{\mathcal{B}_k}$ in \eqref{eq:def:mixedl2l1norm}
is necessary for a fair evaluation of blocks of different sizes (see, e.g., \cite{Yuan:l12,Nardi:BlockSpare,Wei:BlockSparse,Huang:Review} for statistical viewpoints).
Thus, in order to estimate the target signal $\bm{x}^{\star} \in \mathbb{R}^{N}$ accurately,
for every $k \in \{1,\ldots,K\}$,
the scalars
$(x_n^{\star})_{n \in \mathcal{B}_k}$ within the block $\mathcal{B}_k$ should have similar magnitudes.
In many applications, e.g.,
\cite{Yu:Audio,Cevher:Video,Kuroda_FPC,Kitahara:PAWR_GRSS},
it is difficult to set this desired block partition before recovering the signal
because the target signals are block-sparse with unknown block partitions.

\subsection{The LOP-$\ell_2/\ell_1$ penalty and its known properties}
To cope with the unknown block partition, the study in \cite{Kuroda:BlockSparse} 
designs the LOP-$\ell_2/\ell_1$ penalty $\psi_{\alpha}\colon\mathbb{R}^{N}\to\mathbb{R}_{+}$ by
\begin{align}
\label{eq:ProposedConvexLOPpenalty}
\psi_{\alpha}(\bm{x}):=
\min_{\substack{\bm{\sigma} \in \mathbb{R}^{N} \vspace{1.5pt} \\ \|\bm{D}\bm{\sigma} \|_{1} \leq \alpha}} \sum_{n = 1}^{N} \phi (x_{n},\sigma_{n}),
\end{align}
where the block partition is implicitly optimized by the latent variable $\bm{\sigma}$,
the tuning parameter $\alpha \in \mathbb{R}_{+}$ corresponds to the number of blocks,
the proper lower semicontinuous convex function
$\phi\colon\mathbb{R}\times\mathbb{R} \to \mathbb{R}_{+}\cup\{\infty\}$
is defined by
\begin{align}
\label{eq:defenition:phi}
\phi (x_{n},\sigma_{n}) :=
\begin{dcases}
\frac{x_{n}^2}{2\sigma_{n}} + \frac{\sigma_{n}}{2}, &\mbox{{\rm if} }\sigma_{n} > 0;\\
0, &\mbox{{\rm if} } x_{n}=0 \mbox{ {\rm and} } \sigma_{n} = 0;\\
\infty, &\mbox{{\rm otherwise}},
\end{dcases}
\end{align}
and $\bm{D} \in \mathbb{R}^{(N-1) \times N}$ is the first-order difference operator
that maps $\bm{\sigma} \in \mathbb{R}^{N}$ to
$(\sigma_{n+1}-\sigma_{n})_{n=1}^{N-1} \in \mathbb{R}^{N-1}$.
The LOP-$\ell_2/\ell_1$ penalty in
\eqref{eq:ProposedConvexLOPpenalty} is derived as a
tight convex relaxation of an ideal nonconvex penalty
$\psi_{K}^{(\mathrm{nc})}\colon\mathbb{R}^{N}\to\mathbb{R}_{+}$
that takes the minimum of the mixed $\ell_2/\ell_1$ norm over
the possible block partitions
(see \cite[Sect.~II]{Kuroda:BlockSparse} for details).

It remains a major question whether the LOP-$\ell_{2}/\ell_{1}$ penalty $\psi_{\alpha}$
becomes the mixed $\ell_2/\ell_1$ norm using
the desired block partition described in Sect.~\ref{sect:desiredBlockl2l1}.
To date, the performance of $\psi_{\alpha}$ has only been verified experimentally in \cite{Kuroda:BlockSparse,Kuroda:GraphSparse,Kuroda:PSDEstimation,Yu:l0LOP,Furuhashi:LOP_image}.
From a theoretical aspect, the behavior of $\psi_{\alpha}$
has been revealed
only for the extreme choices $\alpha = 0$ and
$\alpha \rightarrow \infty$
i.e., $\psi_{0}(\bm{x}) = \sqrt{N}\|\bm{x} \|_{2} \,\,(= \|\bm{x} \|_{2,1}^{\{1,\ldots,N \}})$
and $\lim_{\alpha \rightarrow \infty}\psi_{\alpha}(\bm{x})  =
\|\bm{x} \|_1 \,\,(= \|\bm{x} \|_{2,1}^{(\{k\})_{k=1}^{N} } )$
for every $\bm{x} \in \mathbb{R}^{N}$
\cite[Theorem 2]{Kuroda:BlockSparse}.
Investigating the behavior of $\psi_{\alpha}$ for general $\alpha \in (0,\infty)$
is a nontrivial question because the minimization in \eqref{eq:ProposedConvexLOPpenalty}
contains the discontinuous function $\phi$ and the nonsmooth function $\|\cdot\|_1$.

\section{Theoretical guarantees for the selection of desired blocks with the LOP-$\ell_2/\ell_1$ penalty}
\label{sect:mainResult}
In this section, we reveal the behavior of the LOP-$\ell_2/\ell_1$ penalty $\psi_{\alpha}$ in \eqref{eq:ProposedConvexLOPpenalty}
for general $\bm{x} \in \mathbb{R}^{N}$ and $\alpha \in \mathbb{R}_{++}$.
Note that the case $\alpha = 0$ is covered by \cite[Theorem 2]{Kuroda:BlockSparse}.
Since it is difficult to directly deal with the constraint form \eqref{eq:ProposedConvexLOPpenalty},
we consider an alternative form
\begin{align}
\label{eq:LOP_EquivalentAddForm}
\minimize_{\bm{\sigma} \in \mathbb{R}^{N}}~\sum_{n = 1}^{N} \phi (x_{n},\sigma_{n}) 
+ \beta_{\alpha}  \|\bm{D}\bm{\sigma} \|_{1}.
\end{align}
For any $\alpha \in \mathbb{R}_{++}$,
using the properties shown in \cite[Appendix C]{Kuroda:BlockSparse}
and the standard result of Lagrange multipliers \cite[Proposition 27.21]{BC:ConvexAnalysis},
we can show that
$\hat{\bm{\sigma}} \in \mathbb{R}^{N}$ is a minimizer 
of the right-hand side of \eqref{eq:ProposedConvexLOPpenalty}
if and only if there exists $\beta_{\alpha} \in \mathbb{R}_{+}$
such that $\hat{\bm{\sigma}}$ is a minimizer of \eqref{eq:LOP_EquivalentAddForm}.

As a main theoretical result of this paper,
we present the next theorem showing that
$\psi_{\alpha}(\bm{x})$ becomes
the mixed $\ell_{2}/\ell_{1}$ norm of $\bm{x} \in \mathbb{R}^{N}$ with
the desired block partition, i.e., blocks
of components of similar magnitudes
(see also Remark \ref{rmrk:IntuitiveInterpretationLOPEqualL2L1} below for an intuitive interpretation of the result).

\begin{theo}
\label{theo:ConditionLOPpenEqualL2L1}
Fix $\bm{x} \in \mathbb{R}^{N}$
and $\alpha \in \mathbb{R}_{++}$
arbitrarily, and set $\beta_{\alpha} \in \mathbb{R}_{+}$
such that the equivalence between \eqref{eq:ProposedConvexLOPpenalty}
and \eqref{eq:LOP_EquivalentAddForm} holds.
Define $\mathcal{J} := \mathrm{supp}(\bm{x})$,
and let
$(\mathcal{B}_k^{\ast})_{k=1}^{K^{*}}$ be a block partition of $\mathcal{J}$, i.e.,
$\bigcup_{k=1}^{K^{*}}\mathcal{B}_k^{*} = \mathcal{J}$,
$\mathcal{B}_k^{*} \neq \varnothing  \,\,  (k = 1,\ldots,K^{*})$,
$\mathcal{B}_k^* \cap \mathcal{B}_{k'}^* = \varnothing  \,\,  (k \neq k')$,
and there exist $(n_k, m_k) \in \{1,\ldots,N \}^2$ such that
$\mathcal{B}_k^{*} = \{n \in \{1,\ldots,N\}\mid n_{k} \leq n \leq m_k \}$
for each $k = 1,\ldots,K^{*}$.
Define the block-diagonal matrix $\bm{U}$ by
\begin{align}
\label{eq:def:OrthogonalBasisNullDiff}
\bm{U} := \begin{pmatrix}
\bm{1}_{|\mathcal{B}^{\ast}_1|}  & &\\
& \ddots &\\
 & &\bm{1}_{|\mathcal{B}^{\ast}_{K^{\ast}}|}
\end{pmatrix}\in \mathbb{R}^{|\mathcal{J}|\times K^{\ast}},
\end{align}
where $\bm{1}_{|\mathcal{B}^{\ast}_k|}$ is the all-one vector of size $|\mathcal{B}^{\ast}_k|$.
Suppose that $\beta_{\alpha} \leq 1/4$,
and
let $\hat{\bm{\sigma}} \in \mathbb{R}^{N}$, $\hat{\bm{\eta}} \in \mathbb{R}^{K^{*}}$,
and $\hat{\mathcal{I}} \subset \{1,\ldots,N-1\}$ satisfy
\begin{align}
\label{eq:expressionOptimalSigmaLOPpenalty}
\hat{\sigma}_{n} &=
\begin{dcases}
0 &\mbox{{\rm if} }  n \notin \mathcal{J};\\
\hat{\varsigma}_1 := \frac{\|\bm{x}_{\mathcal{B}_1^{\ast}} \|_2}{\sqrt{ |\mathcal{B}_1^{\ast}|+ 2\beta_{\alpha}\hat{\eta}_1 } }
&\mbox{{\rm if} }  n \in \mathcal{B}_1^{\ast};\\
\qquad \vdots & \\
\hat{\varsigma}_{K^{\ast}} := \frac{\|\bm{x}_{\mathcal{B}_{K^{\ast}}^{\ast}} \|_2}{\sqrt{ |\mathcal{B}_{K^{\ast}}^{\ast}|+ 2\beta_{\alpha}\hat{\eta}_{K^{\ast}} } }
&\mbox{{\rm if} }  n \in \mathcal{B}_{K^{\ast}}^{\ast},
\end{dcases}\\
\label{eq:def:EtaHat_UtDtSgnDsigma}
\hat{\bm{\eta}} &= \bm{U}^{\mathsf{T}}\bm{D}_{\hat{\mathcal{I}}, \mathcal{J}}^{\mathsf{T}}
(\mathrm{sgn}(\bm{D}\hat{\bm{\sigma}}))_{\hat{\mathcal{I}}},\\
\label{eq:def:SupportDiffSigmaHat}
\hat{\mathcal{I}} &= \mathrm{supp}(\bm{D}\hat{\bm{\sigma}}).
\end{align}
Furthermore, suppose that
\begin{align}
\label{eq:assumpDifferentStdAdjBlk}
&\hat{\varsigma}_k \neq \hat{\varsigma}_{k+1} \quad (k = 1,\ldots,K^{\ast}-1),\\
\label{eq:assumpAbsCompBlkCloseToStd}
&\bigl\|\bigl(\tfrac{|x_{n}|^2}{2\hat{\sigma}_{n}^2} - \tfrac{1}{2}\bigr)_{n\in\mathcal{J}}
-\beta_{\alpha}\bm{D}_{\hat{\mathcal{I}}, \mathcal{J}}^{\mathsf{T}}(\mathrm{sgn}(\bm{D}\hat{\bm{\sigma}}))_{\hat{\mathcal{I}}}\bigr\|_{\infty}
\leq \tfrac{\beta_{\alpha}}{\left\|\bm{D}_{\hat{\mathcal{I}}^{c}, \mathcal{J}}^{\mathsf{T},\dagger}\right\|_{\infty}},
\end{align}
where $\bm{D}_{\hat{\mathcal{I}}^{c}, \mathcal{J}}^{\mathsf{T},\dagger}$ is the Moore-Penrose pseudo-inverse of $\bm{D}_{\hat{\mathcal{I}}^{c}, \mathcal{J}}^{\mathsf{T}}$,
and $\hat{\mathcal{I}}^{c} := \{1,\ldots,N-1\}~\backslash~\hat{\mathcal{I}}$.
Then $\hat{\bm{\sigma}}$
is a minimizer of \eqref{eq:LOP_EquivalentAddForm},
and we have
\begin{align}
\label{eq:LOPl2l1EqualMixedL2L1WithDesiredBlock}
\psi_{\alpha}(\bm{x}) = \sum_{k=1}^{K^{\ast}} w_k^{*} \|\bm{x}_{\mathcal{B}_k^{\ast}} \|_2,
\end{align}
where
$w_k^{*} := \sqrt{ |\mathcal{B}_{k}^{\ast}|+ 2\beta_{\alpha}\hat{\eta}_{k} }/2
+ |\mathcal{B}_{k}^{\ast}|/(2\sqrt{ |\mathcal{B}_{k}^{\ast}|+ 2\beta_{\alpha}\hat{\eta}_{k} })$.
\end{theo}
The proof of this theorem is based on showing that $\hat{\bm{\sigma}}$ given by \eqref{eq:def:EtaHat_UtDtSgnDsigma} satisfies a sufficient condition to be a minimizer of \eqref{eq:LOP_EquivalentAddForm} under the assumptions. This sufficient condition is derived by reformulating the minimality condition that $\bm{0}$ is contained in the subdifferential of the objective function in \eqref{eq:LOP_EquivalentAddForm}
under the additional conditions $\bm{\sigma}_{\mathcal{J}^{c}} = \bm{0}$
and $\mathrm{supp}(\bm{D}_{\mathcal{J}}\bm{\sigma}_{\mathcal{J}}) = \hat{\mathcal{I}}$.
Owing to space limitation, details will be presented in a full paper.

\begin{rmrk}
\label{rmrk:IntuitiveInterpretationLOPEqualL2L1}
We provide an intuitive explanation of
Theorem \ref{theo:ConditionLOPpenEqualL2L1} that
$\psi_{\alpha}(\bm{x})$ becomes
the mixed $\ell_{2}/\ell_{1}$ norm with blocks
of components of similar magnitudes.
More specifically, we demonstrate that
the expression in \eqref{eq:LOPl2l1EqualMixedL2L1WithDesiredBlock}
tends to hold for a \emph{desired} block partition
$(\mathcal{B}_k^{\ast})_{k=1}^{K^{*}}$
in the sense of being compatible with the following criteria:
\begin{enumerate}
\item[i)] The (sample) standard deviations
of $(x_{n})_{n \in \mathcal{B}_k^{*}}$ and $(x_{n})_{n \in \mathcal{B}_{k+1}^{*}}$
are different for each $k = 1,\ldots,K^{*}-1$.
\item[ii)] $|x_{n}|$ for $n \in \mathcal{B}_k^{*}$
are close to the standard deviation of $(x_{n})_{n \in \mathcal{B}_k^{*}}$
for every $k \in \{1,\ldots,K^{*}\}$.
\end{enumerate}
Indeed,
criterion (ii) yields that components in $\mathcal{B}_k^{*}$
have similar magnitudes,
while blocks of similar standard deviations are merged into a single block by criterion (i).

We now demonstrate that \eqref{eq:assumpDifferentStdAdjBlk}
and \eqref{eq:assumpAbsCompBlkCloseToStd} -- the sufficient conditions to derive \eqref{eq:LOPl2l1EqualMixedL2L1WithDesiredBlock} --
tend to hold for $(\mathcal{B}_k^{\ast})_{k=1}^{K^{*}}$ constructed by criteria (i) and (ii).
For simplicity, we assume that $\beta_{\alpha}$ is small enough.
Then, for every $k \in \{1,\ldots,K^{*}\}$, $\hat{\varsigma}_k$
in \eqref{eq:expressionOptimalSigmaLOPpenalty}
is close to the standard deviation 
of $(x_{n})_{n \in \mathcal{B}_k^{*}}$
since $\beta_{\alpha}|\hat{\eta}_k|$ is small enough.
Thus, condition \eqref{eq:assumpDifferentStdAdjBlk}
tends to hold if $(\mathcal{B}_k^{\ast})_{k=1}^{K^{*}}$ is constructed based on criterion (i).
Meanwhile, for small enough $\beta_{\alpha}$,
condition \eqref{eq:assumpAbsCompBlkCloseToStd} tends
to hold if
$\frac{|x_{n}|^2}{2\hat{\sigma}_{n}^2} - \frac{1}{2} \approx 0$,
i.e., $|x_{n}|^2 \approx \hat{\varsigma}_k^2$
for every $n \in \mathcal{B}_k^{*}$ and $k \in \{1,\ldots,K^{*}\}$, which can be expressed as criterion (ii).
\end{rmrk}

\begin{rmrk}
The weight in \eqref{eq:def:mixedl2l1norm}
is justified from statistical viewpoints (see, e.g., \cite{Yuan:l12,Nardi:BlockSpare,Wei:BlockSparse,Huang:Review}
for detail).
For small enough $\beta_\alpha$,
the weight $w_k^{*}$ in \eqref{eq:LOPl2l1EqualMixedL2L1WithDesiredBlock}
is consistent with the proper choice
$
w_k^{*} \approx \sqrt{|\mathcal{B}_k^{*}|}.
$
\end{rmrk}

\begin{rmrk}
One may concern about
the existence of $(\hat{\bm{\sigma}},\hat{\bm{\eta}},\hat{\mathcal{I}})$
satisfying the implicit equations
\eqref{eq:expressionOptimalSigmaLOPpenalty},
\eqref{eq:def:EtaHat_UtDtSgnDsigma}, and \eqref{eq:def:SupportDiffSigmaHat}.
To address this question,
we show that $(\hat{\bm{\sigma}},\hat{\bm{\eta}},\hat{\mathcal{I}})$ satisfying the implicit equations
can be constructed
from given $\mathcal{B}_k^{\ast}$ $(k=1,\ldots,K^{*})$, provided that $\beta_{\alpha}$ is small enough.
First, for $\beta_{\alpha} = 0$, we can compute $\hat{\bm{\sigma}}$
directly by \eqref{eq:expressionOptimalSigmaLOPpenalty}
without $\hat{\bm{\eta}}$ and $\hat{\mathcal{I}}$.
Let $\hat{\bm{\sigma}}_0$ be the vector defined by 
\eqref{eq:expressionOptimalSigmaLOPpenalty} for $\beta_{\alpha} = 0$,
and $\hat{\bm{\eta}}_{0}$ and $\hat{\mathcal{I}}_0$
 given by 
\eqref{eq:def:EtaHat_UtDtSgnDsigma} and \eqref{eq:def:SupportDiffSigmaHat}, respectively, for $\hat{\bm{\sigma}} = \hat{\bm{\sigma}}_0$.
Let $\hat{\bm{\sigma}}'$ be the vector computed by \eqref{eq:expressionOptimalSigmaLOPpenalty} with $\hat{\bm{\eta}} = \hat{\bm{\eta}}_{0}$ and $\beta_\alpha \neq 0$.
If $\beta_\alpha$ is small enough,
then the signs of components of $\bm{D}\hat{\bm{\sigma}}'$ remain unchanged
from $\bm{D}\hat{\bm{\sigma}}_0$,
and hence $(\hat{\bm{\sigma}}',\hat{\bm{\eta}}_{0},\hat{\mathcal{I}}_0)$ satisfy the implicit equations
\eqref{eq:expressionOptimalSigmaLOPpenalty},
\eqref{eq:def:EtaHat_UtDtSgnDsigma}, and \eqref{eq:def:SupportDiffSigmaHat}.
\end{rmrk}

%\begin{rmrk}
%The condition \eqref{eq:assumpAbsCompBlkCloseToStd} suggests that
%we can control the deviation of the magnitudes of components
%in each block by changing $\beta_\alpha$.
%
%Note that $\beta_\alpha$ is related to the tuning parameter $\alpha$
%of the LOP-$\ell_2/\ell_1$ penalty.
%
%
%\begin{align*}
%\bigl\|\bm{D}_{\hat{\mathcal{I}}, \mathcal{J}}^{\mathsf{T}}(\mathrm{sgn}(\bm{D}\hat{\bm{\sigma}}))_{\hat{\mathcal{I}}}\bigr\|_{\infty} \leq 2
%\end{align*}
%since $\bm{D}_{\hat{\mathcal{I}}, \mathcal{J}}^{\mathsf{T}}$ has at most two nonzero entries of magnitude one.
%
%since $|\hat{\eta}_k|$ is bounded above as
%\begin{align*}
%|\hat{\eta}_k| &=
%\bigl|\bm{1}_{|\mathcal{B}^{\ast}_k|}^{\mathsf{T}}\bm{D}_{\hat{\mathcal{I}}, \mathcal{J}}^{\mathsf{T}}(\mathrm{sgn}(\bm{D}\hat{\bm{\sigma}}))_{\hat{\mathcal{I}}}\bigr|
%\leq |\mathcal{B}_k^{\ast}| \|\bm{D}_{\hat{\mathcal{I}}, \mathcal{J}}^{\mathsf{T}}(\mathrm{sgn}(\bm{D}\hat{\bm{\sigma}}))_{\hat{\mathcal{I}}}\|_{\infty}\\
%&\leq |\mathcal{B}_k^{\ast}|\|\bm{D}_{\hat{\mathcal{I}}, \mathcal{J}}^{\mathsf{T}}\|_{\infty}
%\|(\mathrm{sgn}(\bm{D}\hat{\bm{\sigma}}))_{\hat{\mathcal{I}}} \|_{\infty}
%\leq|\mathcal{B}_k^{\ast}|\|\bm{D}^{\mathsf{T}}\|_{\infty} = 2 |\mathcal{B}_k^{\ast}|.
%\end{align*}
%\end{rmrk}

\setlength{\abovedisplayskip}{5pt}
\setlength{\belowdisplayskip}{5pt}

\section{Application of the LOP-$\ell_2/\ell_1$ penalty for APS Estimation in MIMO systems}
\subsection{Problem formulation}
We show that the APS estimation can be formulated as a linear inverse problem using
a sample channel covariance matrix in MIMO wireless communications.
For simplicity, we consider the uplink of a MIMO wireless channel
between a single-antenna user and a base station equipped with $M$
antennas in a 2D, i.e., azimuth only, scenario (see, e.g., \cite{Miretti:CovarianceConversion} for more general scenarios).
Let $\bm{R} \in \mathbb{C}^{M \times M}$ be a channel covariance matrix, which
can be modeled by
\begin{align}
\label{eq:FormulaBetweenAPSandTrueChannelCovarianceMatrix}
\bm{R} = \mbox{$\sum_{n=1}^{N}$} x_{n}^{\star}\bm{a}(\theta_{n}) \bm{a}(\theta_{n})^{\mathsf{H}}
+ \bm{\mathcal{E}},
\end{align}
where
$\bm{x}^{\star} := (x_1^{\star},\ldots,x_{n}^{\star})^{\mathsf{T}} \in \mathbb{R}_{+}^{N}$
is the APS on the discrete grid
$\theta_1, \ldots, \theta_{N}$ of angles
in $\Omega \subset [-\pi,\pi)$,
$\bm{a}(\theta) \in \mathbb{C}^{M}$ is the array response vector
for $\theta \in \Omega$,
and $\bm{\mathcal{E}} \in \mathbb{C}^{M \times M}$
is an error term due to modeling and discretization.
We further assume that the estimate $\hat{\bm{h}}_t \in \mathbb{C}^{M}$
of the channel at the $t \in \{1,\ldots,T\}$ coherence block is given by
$\hat{\bm{h}}_t = \bm{h}_t + \bm{n}_t$,
where the channel $\bm{h}_t \in \mathbb{C}^{M}$ follows the complex Gaussian distribution $\mathcal{CN}(\bm{0},\bm{R})$,
and $\bm{n}_t \in \mathbb{C}^{M}$ is the complex white Gaussian noise of variance $s^2$.
Using $\hat{\bm{h}}_1,\ldots,\hat{\bm{h}}_{T}$,
we can compute an estimate of $\bm{R}$ by
\begin{align}
\label{eq:def:EstimateChannelCovarianceMatrixWithStructure}
\hat{\bm{R}} =
P_{\mathcal{C}}\Bigl(\tfrac{1}{T}\mbox{$\sum_{t=1}^{T}$}\hat{\bm{h}}_t \hat{\bm{h}}_t^{\mathsf{H}}  - s^2 \bm{I}\Bigr),
\end{align}
where $\mathcal{C} \subset \mathbb{C}^{M \times M}$ is a constraint set to exploit the structure of the covariance matrix.
Combining \eqref{eq:FormulaBetweenAPSandTrueChannelCovarianceMatrix}
and \eqref{eq:def:EstimateChannelCovarianceMatrixWithStructure},
we have
\begin{align}
\label{eq:MatFormObservationModelAPS}
\hat{\bm{R}} = \mbox{$\sum_{n=1}^{N}$} x_{n}^{\star}\bm{a}(\theta_{n}) \bm{a}(\theta_{n})^{\mathsf{H}}
+ \tilde{\bm{\mathcal{E}}},
\end{align}
where $\tilde{\bm{\mathcal{E}}} \in \mathbb{C}^{M \times M}$
is an error term expected to have a small enough norm.
Note that
$\tilde{\bm{\mathcal{E}}} \in \mathbb{C}^{M \times M}$
includes the estimation error in \eqref{eq:def:EstimateChannelCovarianceMatrixWithStructure}
and also the modeling and discretization errors in \eqref{eq:FormulaBetweenAPSandTrueChannelCovarianceMatrix}.
Since \eqref{eq:MatFormObservationModelAPS} usually contains 
redundant equations because of the array model (see Example \ref{exmp:UniformLinearArrayDirective} below),
by extracting essential $\bar{M}$ equations,
we obtain a concise observation model
\begin{align}
\label{eq:ObservationModelForAPS}
\hat{\bm{r}} = \bm{A}\bm{x}^{\star} + \tilde{\bm{\varepsilon}},
\end{align}
where
$\hat{\bm{r}} \in \mathbb{R}^{\bar{M}}$, $\bm{A}\in \mathbb{R}^{\bar{M}\times N}$,
and $\tilde{\bm{\varepsilon}} \in \mathbb{R}^{\bar{M}}$
are extracted from $\hat{\bm{R}}$, $\bm{a}(\theta_{n}) \bm{a}(\theta_{n})^{\mathsf{H}}$,
and $\tilde{\bm{\mathcal{E}}}$, respectively.
An example of the array model
and the corresponding constraint set $\mathcal{C}$ in \eqref{eq:def:EstimateChannelCovarianceMatrixWithStructure} are provided below.

\begin{exmp}
\label{exmp:UniformLinearArrayDirective}
In the 3GPP document \cite[Section 7.1]{TechRep:3gpp},
an array response vector is modeled by
$
\bm{a}(\theta) =
g(\theta)
[1, e^{\imath 2 \pi \frac{d}{\lambda_{\mathrm{c}}}\sin (\theta)}, \ldots,e^{\imath 2 \pi \frac{d}{\lambda_{\mathrm{c}}}(N-1)\sin (\theta)}]^{\mathsf{T}}/\sqrt{N}
$
with the directive gain
$
g(\theta) = -\min \{12\left(\theta/\theta_{\mathrm{3db}})^2, \mathrm{SLA} \right\} \, [\mathrm{dB}]$
for $\theta \subset [-\pi/2, \pi/2]$,
where $\theta_{\mathrm{3dB}} = \ang{65}$,
$\mathrm{SLA} = 30 \, [\mathrm{dB}]$,
$\lambda_{\mathrm{c}} \in \mathbb{R}_{++}$ is the carrier wavelength, and
$d \in \mathbb{R}_{++}$ is the antenna spacing.
Since $\bm{a}(\theta) \bm{a}(\theta)^{\mathsf{H}}$ is Toeplitz
for any $\theta \in \Omega$,
$
\sum_{n=1}^{N} x_{n}^{\star}\bm{a}(\theta_{n}) \bm{a}(\theta_{n})^{\mathsf{H}}
$
is also Toeplitz
for any $\bm{x}^{\star} \in \mathbb{R}^{N}$.
Thus, we set
$\mathcal{C} = \mathcal{T} \cap \mathcal{S}_{+}$ in \eqref{eq:def:EstimateChannelCovarianceMatrixWithStructure},
where $\mathcal{T}$ and $\mathcal{S}_{+}$
are the sets of all Toeplitz matrices and all positive semi-definite matrices, respectively.
Since $\hat{\bm{R}}$ and $\bm{a}(\theta_{n}) \bm{a}(\theta_{n})^{\mathsf{H}}$
are Hermitian and Toeplitz,
we have at most $2M$ different equations in
\eqref{eq:ObservationModelForAPS}.
\end{exmp}

Following the studies in \cite{Song:ML_APS_GroupSparse,Khalilsarai:ML_APS_GroupSparse,Cavalcante:ModelDataAPS}, we also assume that a dataset of past APS, denoted by
$\mathcal{M} := \{\bar{\bm{x}}_{1},\ldots,\bar{\bm{x}}_{L}\} \subset \mathbb{R}_{+}^{N}$, is available.
Thus, our goal is to estimate the true APS $\bm{x}^{\star}$
from the measurements $(\hat{\bm{r}},\bm{A})$ in \eqref{eq:ObservationModelForAPS} and the dataset $\mathcal{M}$.

In MIMO systems, when the antenna array is sufficiently large and the
channel covariance matrices are reliably estimated, even simple linear
processing techniques can provably deliver good performance for APS
estimation \cite{renato18error}.
However, APS estimation becomes challenging with
moderate or small arrays and/or when covariance estimates are
inaccurate owing to finite-sample effects \cite{Kaneko:APS_AdaptGaussKernel,Cavalcante:PilotDecont}. APS estimation in this
small-array regime is especially relevant for applications beyond
channel conversion, e.g., deep-learning-based localization \cite{Decurninge:LocalizationMIMO}
in private networks or on the user terminal side, where the number $M$ of
antennas is typically much smaller than at base stations in public
networks.
The number $L$ of past data is also typically small because of
the variability of the propagation environment.
In such situations, it is difficult to obtain satisfactory estimation accuracy
with algorithms based only on the physical observation model and past data.

\setlength{\abovedisplayskip}{4pt}
\setlength{\belowdisplayskip}{4pt}

\subsection{Proposed method for APS estimation}
To tackle the challenging APS estimation problem,
we propose to exploit block-sparsity of APS
by the LOP-$\ell_2/\ell_1$ penalty.
As indicated in \cite{Song:ML_APS_GroupSparse,Khalilsarai:ML_APS_GroupSparse},
APS is block-sparse in many practical propagation environments.
However, the block-sparsity of APS has been difficult to exploit because
the desired block partition depends on the unknown propagation environment.
Motivated by our theoretical result (Theorem \ref{theo:ConditionLOPpenEqualL2L1}),
we apply the LOP-$\ell_2/\ell_1$ penalty for addressing this difficulty.

We also propose to use
the generalized Moreau enhanced (GME) version of the LOP-$\ell_2/\ell_1$ penalty.
While Theorem \ref{theo:ConditionLOPpenEqualL2L1} demonstrates that the LOP-$\ell_2/\ell_1$ penalty
is free from the underestimation caused by the block containing components of widely different magnitudes,
the LOP-$\ell_2/\ell_1$ penalty tends to underestimate large-magnitude components because
its value increases with the magnitudes of nonzero components in $\bm{x}$
(see \eqref{eq:LOPl2l1EqualMixedL2L1WithDesiredBlock}).
To remedy this drawback, following the approach of \cite{Kuroda:CNC_GME_MI}, we design the GME-LOP-$\ell_2/\ell_1$ penalty
$\Psi_{\bm{B},\alpha}\colon\mathbb{R}^{N}\rightarrow\mathbb{R}$
as the difference between the LOP-$\ell_2/\ell_1$ penalty and its generalized Moreau envelope, i.e.,
\begin{align*}
\Psi_{\bm{B},\alpha}(\bm{x}) :=  \psi_{\alpha}(\bm{x}) - \min_{\bm{v} \in \mathbb{R}^{N}}\left[\psi_{\alpha}(\bm{v}) + \tfrac{1}{2}\|\bm{B}(\bm{x}-\bm{v})\|_2^2\right],
\end{align*}
where $\bm{B} \in \mathbb{R}^{J \times N}$ can be adjusted to control the shape of $\Psi_{\bm{B},\alpha}$.
The next proposition on the boundedness of $\Psi_{\bm{B},\alpha}$
suggests that
the generalized Moreau enhancement
remedies the increase of the penalty value, and hence mitigates the underestimation tendency.

\begin{prop}
\label{prop:boundedGME}
Suppose that $\bm{B}^{\mathsf{T}}\bm{B}$ is nonsingular.
Then $\Psi_{\bm{B},\alpha}$ is a bounded function for any $\alpha \in \mathbb{R}_{+}$.
\end{prop}

To simultaneously exploit the block-sparsity,
the physical observation model, and the dataset of past APS,
we design the proposed optimization problem:
\begin{align}
\label{eq:def:ProposedModelForAPSestimation}
\minimize_{\bm{x} \in \mathbb{R}_{+}^{M}} \tfrac{1}{2}\|\bm{A}\bm{x}-\hat{\bm{r}}\|_2^2
+\tfrac{\mu}{2}\|\bm{x}-\bar{\bm{x}}\|_{\bm{P}}^2 +\lambda \Psi_{\bm{B},\alpha}(\bm{x}),
\end{align}
where $\mu \in \mathbb{R}_{++}$ and $\lambda \in \mathbb{R}_{++}$
are tuning parameters to control the importance of the dataset and the block-sparsity, respectively.
The first term in \eqref{eq:def:ProposedModelForAPSestimation} measures the error for the observation model \eqref{eq:ObservationModelForAPS}.
The second term in \eqref{eq:def:ProposedModelForAPSestimation}
is designed based on the idea in \cite{Cavalcante:ModelDataAPS}.
It measures the distance between
$\bm{x}$ and the distribution of the dataset by setting
$\bar{\bm{x}} := \frac{1}{L}\sum_{l=1}^{L}\bar{\bm{x}}_{l}$,
$
\bm{C} := \frac{1}{L-1}\sum_{l=1}^{L}
(\bar{\bm{x}}_l - \bar{\bm{x}})(\bar{\bm{x}}_{l} - \bar{\bm{x}})^{\mathsf{T}}
$,
$
\bm{P} := (\bm{C} + \delta \bm{I})^{-1}
$ with a small value $\delta > 0$ to avoid numerical instability,
and $(\forall \bm{z} \in \mathbb{R}^{N})\,\,\|\bm{z}\|_{\bm{P}} := \sqrt{\bm{z}^{\mathsf{T}}\bm{Pz}}$.

Based on the result in \cite{Kuroda:CNC_GME_MI}, we can
guarantee convexity of
problem \eqref{eq:def:ProposedModelForAPSestimation} by setting
$\bm{B}^{\mathsf{T}}\bm{B} = (\omega/\lambda)(\bm{A}^{\mathsf{T}}\bm{A} + \mu\bm{P})$
with any $\omega \in [0,1]$.
Note that the assumption of Proposition \ref{prop:boundedGME}
is also satisfied if $\omega \in (0,1]$.
We can iteratively compute a global minimizer of problem \eqref{eq:def:ProposedModelForAPSestimation}
under its convexity condition
with a slightly modified algorithm of \cite{Kuroda:CNC_GME_MI}.

\section{Simulations}
We present numerical simulations for APS estimation in MIMO systems
to investigate how the estimation accuracy is improved with the proposed use of block-sparsity in \eqref{eq:def:ProposedModelForAPSestimation}.
We consider the
uniform linear array with directive antennas presented in Example \ref{exmp:UniformLinearArrayDirective}
based on the 3GPP document \cite[Section 7.1]{TechRep:3gpp}.
The parameters in Example \ref{exmp:UniformLinearArrayDirective}
are set as follows.
The angles
$\theta_1,\ldots,\theta_{N}$
are uniformly chosen from $\Omega = [-\pi/2, \pi/2]$ with $N=100$,
the carrier wavelength is computed by $\lambda_{\mathrm{c}} = 3\cdot 10^{8}/f_{\mathrm{c}}$
with the carrier frequency $f_{\mathrm{c}} = 2.1 \,\, [\mathrm{GHz}]$,
and the antenna spacing $d$ is set to $\lambda_{\mathrm{c}}/2$.
The number $M$ of antennas are chosen from $\{4,8,12,\ldots,32\}$.

Following the previous studies in \cite{Cavalcante:PilotDecont,Miretti:CovarianceConversion,Bameri:CovarianceConversion,Kaneko:APS_AdaptGaussKernel}, we generate the true APS
$x_{n}^{\star} = \sum_{q=1}^{Q}a_{q} f_{q}(\theta_n)$
for $n = 1,\ldots,N$,
where $Q$ is uniformly drawn from $\{1,2\}$,
$a_{q}$ is uniformly drawn from $[0,1]$
and further normalized to satisfy $\sum_{q=1}^{Q} a_q = 1$,
and
$f_{q}(\theta) := \mathcal{N}(\phi_{q},\Delta_{q}^{2})$
with $\phi_{q}$ and $\Delta_{q}$
uniformly drawn from $[-2\pi/5,-\pi/5]$
and $[\ang{2}, \ang{4}]$, respectively.
For each run of the simulations,
we define the channel covariance matrix according to
\eqref{eq:FormulaBetweenAPSandTrueChannelCovarianceMatrix}
using the generated APS,
where we set $\bm{\mathcal{E}} = \bm{O}$ for simplicity.
Then we compute the sample channel covariance matrix using
\eqref{eq:def:EstimateChannelCovarianceMatrixWithStructure},
where the number $T$ of channel samples is set to $1000$,
and the noise variance $s^2$ is determined so that the SNR
$E[\|\bm{h}_t\|^2]/N/s^2$ becomes $30 \,\mathrm{dB}$,
where $E[\|\bm{h}_t\|^2]$ is approximated
using the samples $\bm{h}_1,\ldots,\bm{h}_{T}$.
We approximate $P_{\mathcal{C}}$ in \eqref{eq:def:EstimateChannelCovarianceMatrixWithStructure}
with $1000$ iterations of Halpern's algorithm \cite[Sect.~30.1]{BC:ConvexAnalysis}
using the projections $P_{\mathcal{T}}$ and $P_{\mathcal{S}_{+}}$
(see also Example \ref{exmp:UniformLinearArrayDirective}).
To consider the deviation of the true APS and the dataset,
we generate
APS in the dataset $\mathcal{M}$ 
with $Q$ and $\phi_{q}$
uniformly drawn from $\{1,2,3,4,5\}$ and
$[-2\pi/5,2\pi/5]$, respectively, with the other settings identical to those used in the generation of the 
true APS.

To see how the performance is improved with the proposed use of block-sparsity,
we compare the proposed method with the projected gradient algorithm
solving problem \eqref{eq:def:ProposedModelForAPSestimation} with $\lambda = 0$.
Note that this projected gradient algorithm is essentially the hybrid model-data driven algorithm of \cite{Cavalcante:ModelDataAPS}
since the solution to \eqref{eq:def:ProposedModelForAPSestimation} is unique if $\mu > 0$.
For problem \eqref{eq:def:ProposedModelForAPSestimation},
we compute
the mean $\bar{\bm{x}}$ and covariance matrix $\bm{C}$ of
past APSs
using $L=1000$ samples,
and then we compute $\bm{P}$ in \eqref{eq:def:ProposedModelForAPSestimation}
with $\delta = \|\bm{C}\|_{2}/100$.
For reference, we also show
the off-the-shelf solver ($\mathtt{lsqnonneg}$ in MATLAB)
for the nonnegative least squares (NNLS) problem
$\minimize_{\bm{x} \in \mathbb{R}_{+}^{M}} \tfrac{1}{2}\|\bm{A}\bm{x}-\bm{r}\|_2^2$, which is employed in, e.g., \cite{Khalilsarai:NNLS_APS}.

\begin{figure}[t]
  \centering
    \includegraphics[width=0.68\columnwidth]{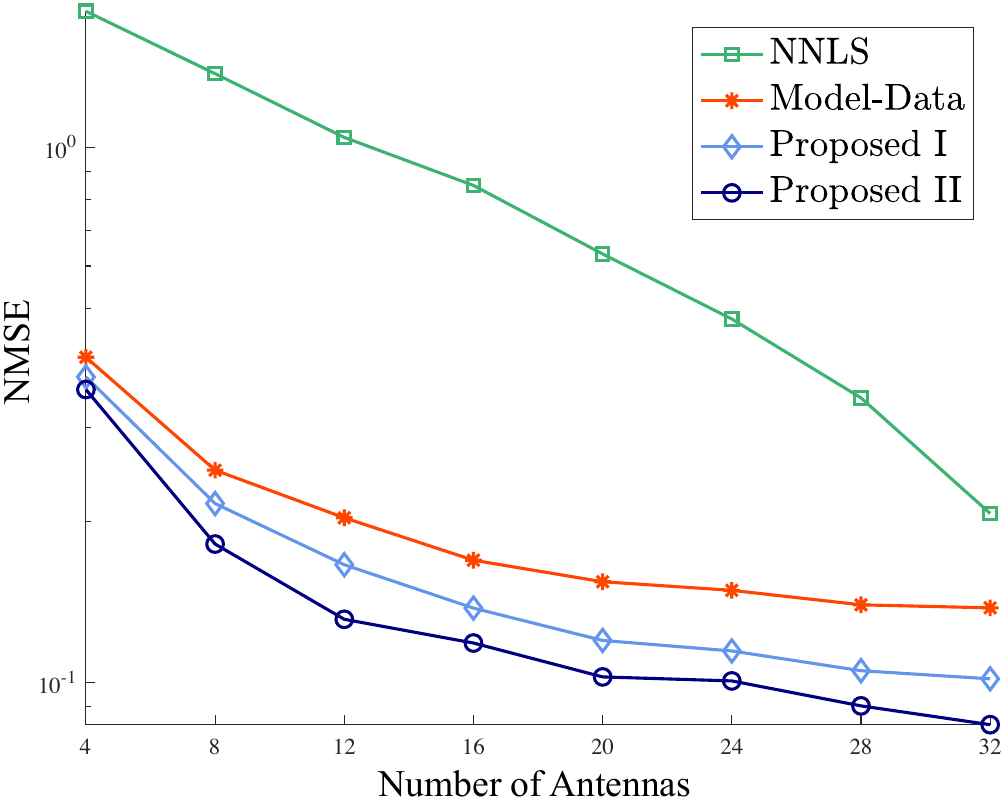}\vspace{-8pt}
  \caption{NMSE versus the number of antennas.}\vspace{-12pt}
  \label{fig:Simulation_APS_MIMO}
\end{figure}

Fig.~\ref{fig:Simulation_APS_MIMO} shows
the normalized mean square error (NMSE)
$\|\bm{x}^{\star}-\hat{\bm{x}}\|_2^2/\|\bm{x}^{\star}\|_2^2$
averaged over $500$ independent trials,
where $\hat{\bm{x}}$ is the estimate of each algorithm, and
``Proposed I''
and ``Proposed II'' use the LOP-$\ell_2/\ell_1$ penalty ($\bm{B} = \bm{O}$ in \eqref{eq:def:ProposedModelForAPSestimation})
and the GME-LOP-$\ell_2/\ell_1$ penalty, respectively.
Tuning parameters of each method are adjusted to obtain the best NMSE for each number of antennas.
NNLS has the worst performance, which can be partially explained by the fact that the block-sparsity
of APS is not exploited.
In contrast, the proposed use of block-sparsity significantly improves the estimation accuracy
from the hybrid model-data driven method.

\fontsize{9pt}{10.5pt}\selectfont
\setlength{\baselineskip}{9pt}
\bibliographystyle{IEEEtran}

\end{document}